\documentclass [aps,prb,twocolumn,unsortedaddress,showkeys,showpacs,
a4paper,unsortedaddress,amsmath,amssymb,byrevtex]{revtex4}

\usepackage[latin1]{inputenc}
\usepackage{graphicx}
\usepackage{dcolumn}
\usepackage{bm}
\usepackage{hyperref}
\usepackage{times}

\begin{document}

\title{Structure and electronic properties of molybdenum monoatomic
wires encapsulated in carbon nanotubes}

\author{A. Garc\'ia-Fuente$^1$}
\author{V. M. Garc\'ia-Su\'arez$^{2,3}$}
\author{J. Ferrer$^{2,3}$}
\author{A. Vega$^1$}

\affiliation{$^1$ Departamento de F\a'{\i}sica Te\a'orica,
At\a'omica y \'Optica. Universidad de Valladolid, E-47011
Valladolid, Spain} \affiliation{$^2$ Departamento de F\a'{\i}sica
\& CINN, Universidad de Oviedo, 33007 Oviedo Spain}
\affiliation{$^3$Department of Physics, Lancaster University,
Lancaster, United Kingdom}
\date{\today}

\begin{abstract}
Monoatomic chains of molybdenum encapsulated in single walled
carbon nanotubes of different chiralities are investigated using
density functional theory. We determine the optimal size of the
carbon nanotube for encapsulating a single atomic wire, as well as
the most stable atomic arrangement adopted by the wire. We also
study the transport properties in the ballistic regime by computing
the transmission coefficients and tracing them back to electronic
conduction channels of the wire and the host. We predict that
carbon nanotubes of appropriate radii encapsulating a Mo wire have
metallic behavior, even if both the nanotube and the wire are
insulators. Therefore, encapsulating Mo wires in CNT is a way to
create conductive quasi one-dimensional hybrid nanostructures.
\end{abstract}



\maketitle

\section{Introduction}

Miniaturization is one of the most important goals in modern
electronics. Besides the gain achieved with the reduction of size
and energy consumption in general, approaching the scale of
nanometers also means that new and exciting properties arise due
to quantum confinement effects. At this size regime in which it is
possible to reduce dimensionality and even one atom more or less
can make a difference, the geometrical and electronic structures
and related properties like magnetism, transport, optics,
hardness, etc. are, in general, different to those of the
macroscopic material. Thus, the seek for low-dimensional systems
with potential technological applications is a matter of intense
research in nanoelectronics and spintronics.

\begin{table*}
\caption{Insertion energy of the ground state of the Mo wire encapsulated inside
zigzag and armchair nanotubes of different radii.}
\vspace{10pt}
\begin{tabular}{|c|c|c|c|c|c|}\hline
\multicolumn{3}{|c|}{zigzag nanotubes} & \multicolumn{3}{|c|}{armchair nanotubes} \\
\hline
\, chirality \, & \, radius (\AA) \, & \, Insertion Energy (eV) \, & \, chirality \, & \, radius (\AA) \, & \, Insertion Energy (eV) \,  \\
\hline
(7,0) & 2.82 & -1.775 & (4,4) & 2.79 & -1.355  \\
(8,0) & 3.21 & -2.566 & (5,5) & 3.46 & -2.751  \\
(9,0) & 3.60 & -2.789 & (6,6) & 4.14 & -2.771  \\
(10,0) & 3.98 & -2.791 & (7,7) & 4.81 & -2.544  \\
(11,0) & 4.38 & -2.642 & (8,8) & 5.49 & -2.447  \\
(12,0) & 4.77 & -2.530 &   &   &    \\
(13,0) & 5.16 & -2.472 &   &   &    \\
\hline
\end{tabular}
\label{TI}
\end{table*}

Low-dimensional transition-metal systems have been extensively
investigated due to their interesting electronic and magnetic
properties coming mainly from the geometry-dependent unfilled $d$
band and $sp$-$d$ hybridization effects.\cite {demangeat} For
instance, transition-metal surfaces and interfaces are
two-dimensional systems whose growth and properties are relatively
well controlled and characterized at present.
\cite{spanjaard,bader,somorjai} The ultimate goal however in
reducing the dimensionality in transition-metal structures is the
monoatomic wire, which can be considered an ideal system to
investigate quantum effects. Thus the fabrication of stable
transition-metal monoatomic wires is a challenge that is being
extensively researched at the moment.

Nanotubes are natural quasi one-dimensional nanostructures with
promising applications as building blocks in
nanoelectronics.\cite{baughman} There is a wide variety of
nanotubes depending on the constituent elements, wall size and
chirality. Since they provide space available for encapsulating
different kinds of nanostructures, nanotubes have opened new
prospects for the design of quasi one-dimensional hybrid
nanosystems by filling them with such nanostructures (cluster
arrays, wires).\cite{monthioux} An important aspect  of the
encapsulation into carbon nanotubes, which has technological
implications, is that it prevents the wires inside from
degradation (like oxidation) at ambient conditions. Carbon
nanotubes (CNT) have already been filled with transition
metals\cite{guerret,seraphin,chu,setlur,lcqin,grobert,qin1,qin2,
muramatsu,tao} and other elements like germanium,\cite{loiseau}
iodine\cite{guan1} or lanthanum.\cite{guan2} Several reliable
techniques have been developed to synthesize such quasi
one-dimensional nanostructures inside carbon nanotubes, such as
annealing metal nanowires confined in nanotubes based on the
Rayleigh instability (instability of liquid cylinders due to
surface tension), using fullerene cages and catalytic
decomposition of precursors, using arc-discharged procedures, etc.
In most cases however, either one-dimensional arrays of clusters
or wires with transversal section larger than a single atom were
produced, although Guan {\it et al.} managed to generate a single
atomic chain of iodine longer than 10 nm.\cite{guan1} In the
particular case of Mo, the group of Dresselhaus has done an interesting study.
In their first experimental work \cite{muramatsu} the authors
concluded that monoatomic wires were formed inside CNTs. In their second
work \cite{meunier}, calculations were performed for finite transversal section
Mo bcc wires, and the experimental data were reinterpreted concluding that the
encapsulated wires had such finite transversal section instead of being
monoatomic. Later, in a most recent experimental work,\cite{tao} they have shown
that the arrangements of encapsulated Mo can be controlled to some extend by varying the reaction conditions. Whether monoatomic Mo wires can be encapsulated
or not in CNTs remains still an open question.


From the theoretical point of view, there are still few ab-initio
studies of these types of quasi-one dimensional hybrid systems.
Both Yang {\it et al.}\cite{yang} and Ivanovskaya {\it et
al.}\cite{iva} have studied within density functional theory (DFT)
\cite{kohn-sham} different transition metal nanowires (Ti, Fe, Co,
Zr) inside carbon nanotubes with the aim of exploring their
structural arrangement, electronic properties and magnetic
behavior. Tao et al.\cite{tao} investigated finite transversal section
Mo wires encapsulated in CNTs.

In a recent work\cite{amador} we investigated with the codes
SIESTA \cite{siesta} and SMEAGOL \cite{smeagol} the
electronic structure and transport properties of free-standing
single atomic chains of molybdenum. The ground state of the
periodic monoatomic Mo wire was found to be made of tightly bonded
dimers which were non-magnetic. The formation of the dimers was
due to the strong covalent bond made between pairs of Mo atoms,
necessary to achieve a closed-shell electronic configuration (the
Mo atom has exactly a half filled valence configuration). We found
that such dimerization of the molybdenum atomic wires has dramatic
effects on their electronic and transport properties. While
equidistant wires were found to be metallic and have a very high
zero-bias conductance, dimerized wires showed a large gap which
made them insulators. We hence found that these chains show a
metal-to-insulator transition as a function of the intra-dimer
distance.

Although long single atomic molybdenum wires can not be grown in a
free-standing environment, they could be synthesized in the inner
hollow of a carbon nanotube, which would stabilize them. It is
therefore important to survey this possibility as well as to
analyze the electronic behavior and transport properties of the
resulting one-dimensional hybrid nanostructures. This is the aim
of the present work, in which we have investigated monoatomic
chains of molybdenum encapsulated in single walled carbon
nanotubes of different chiralities. We have determined the optimal
size of the carbon nanotube for encapsulating a single atomic
wire, as well as the most stable atomic arrangement adopted by the
wire inside. We have obtained insight on the bonding between the
Mo chain and its host by analyzing the band-structures and
densities of states. Finally, we have studied the transport
properties in the ohmic regime by computing the transmission
coefficients and tracing them back to electronic conduction
channels of the wire and the host.

The rest of the paper is organized as follows. In next section we
give the details of our theoretical approach. In Section III we
present and discuss the structural properties of encapsulated
wires. Electronic and transport properties are presented in
section IV. The main conclusions are summarized at the end.

\section{Details of the DFT approach}

Our calculations of the electronic structure were performed with
the DFT code SIESTA.\cite{siesta} We calculated the exchange and
correlation potential with the generalized gradient approximation
(GGA) as parametrized by Perdew, Burke and Ernzerhof \cite{PBE}
and the additional non-collinear (NCL) GGA implementation by
Garc\'ia-Su\'arez {\it et al.}\cite{Victor1} We replaced the
atomic cores by nonlocal norm-conserving
Troullier-Martins\cite{TM_1991} pseudopotentials, which were
factorized in the Kleinman-Bilander form\cite{KB_1982} and
generated using the atomic configuration $4d^{5}$ $5s^1$ $5p^0$
with cutoff radii of $1.67$, $2.30$ and $2.46$ a.u., respectively,
in the case of Mo, and $2s^2$ $2p^2$ with cutoff radii of $1.25$
and $1.25$, respectively, in the case of C. This electronic
configuration for Mo was the same as that used by Zhang {\it et
al.}\cite{zhang} in their study of Mo clusters. These authors
concluded that the use of the semicore $4p^6$ states worsens the
results for bulk Mo structure and that the $4d^{5}$ $5s^1$ $5p^0$
configuration is a better choice for clusters larger than the
dimer. We also included nonlinear core corrections for Mo,
generated with a radius of 1.2 a.u, to account for the overlap of
the core charge with the valence $d$ orbitals and to avoid spikes
which often appear close to the nucleus when the GGA approximation
is used. We tested that this pseudopotential reproduced accurately
the eigenvalues of different excited states of the isolated Mo
atom. The pseudopotential of C is the same as the one used by
Garc\'ia-Su\'arez {\it et al.} in their study of
nanotube-encapsulated metallocene chains.\cite{Victor2,Victor3}

SIESTA employs a linear combination of pseudoatomic orbitals to
describe the valence states. The basis set of Mo included
double-$\zeta$ polarized (DZP) orbitals, i.e. two radial functions
to describe the $5s$ shell and another two for each $d$ state of
the $4d$ shell, plus a single radial function for each $p$ state
of the empty $p$ shell. For C, we used a double-$\zeta$ (DZ) basis
with two radial functions to describe the $2s$ shell and another
two for each $p$ state of the $2p$ shell. SIESTA also uses a
numerical grid to compute the exchange and correlation potential,
and to perform the real-space integrals that yield the Hamiltonian
and overlap matrix elements. We defined such grid with an energy
cutoff of 200 Rydberg. We also used 100 $k$ points along the
direction of the nanotube, which were found to be enough to
converge the energy of the system and the band structure.

We also smoothed the Fermi distribution function that enters the
calculation of the density matrix with an electronic temperature
of 300 K and used a conjugate gradient algorithm,\cite{NR} to
relax the atomic positions until the interatomic forces were
smaller than 0.005 eV/\AA. Finally, we performed careful tests for
particular cases to ensure the quality and stability of the basis
set and the real space energy cutoff employed. We found that the
results were hardly modified when the DZP basis of Mo was replaced
by a triple-$\zeta$ doubly polarized basis set. The basis set used
for C has been validated in previous works.\cite{Victor2,Victor3}
Similar results were also obtained by considering an electronic
temperature of 100 K.

We used SMEAGOL\cite{smeagol} to compute the conductance
in the ohmic regime. This code is a flexible and efficient
implementation of the non-equilibrium Green's functions formalism
(NEGF) and is specially designed to calculate the transport
properties of nanoscale systems. SMEAGOL obtains the Hamiltonian
from SIESTA and calculates the density and the transmission with
the NEGF. Using the Landauer formula\cite{datta} and expanding the
transmission coefficients $T(E,V)$ at low energy and low voltage,
the low-voltage differential conductance can be shown to be equal
to

\begin{equation}
G(V)=\frac{\mathrm{d}I}{\mathrm{d}V}\simeq \mathrm{G}_0
T(E_\mathrm{F},0)
\end{equation}

\noindent where $T(E_\mathrm{F},0)$ stands for the zero-voltage
transmission evaluated at the Fermi level, and
$\mathrm{G}_0=2\,\mathrm{e}^2/\mathrm{h}$ is the conductance
quantum. Therefore, the zero-voltage transmission evaluated at
$E_\mathrm{F}$ provides an estimate of the differential
conductance in the ohmic regime. Notice that in a ballistic
one-dimensional periodic system the zero-voltage transmission
coefficient at a given energy just counts the number of bands at
this energy, i.e. the number of transmission channels.

\begin{figure}
\includegraphics[width=0.90\columnwidth]{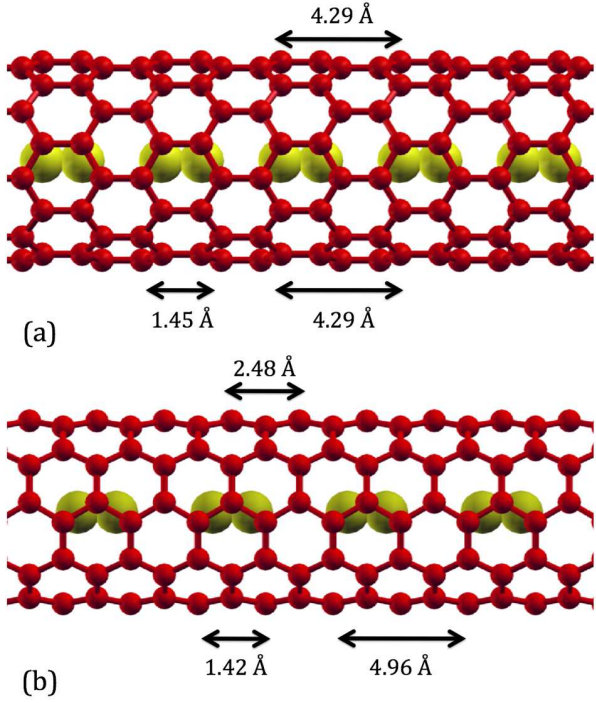}
\caption{\label{Figure1}(Color online) Ground state of a Mo wire
inside (a) a (9,0) zigzag nanotube and (b) a (5,5) armchair
nanotube.}
\end{figure}

\section{Structural Properties}

The ground state of a periodic monoatomic Mo wire in a
free-standing configuration is formed by tightly bonded dimers
with a separation between atoms of 1.58 \AA~and an inter-dimer
distance of 3.01 \AA, which leads to a lattice constant of 4.59
\AA.\cite{amador} The dimer formation comes from the exact half
filling of the Mo electronic shells. Therefore, strong covalent
bonds are made and the resulting wire is non-magnetic. The
free-standing chains have an energy gap of about 1.5 eV around the
Fermi level, which makes them insulators.

\begin{figure}
\includegraphics[width=0.90\columnwidth]{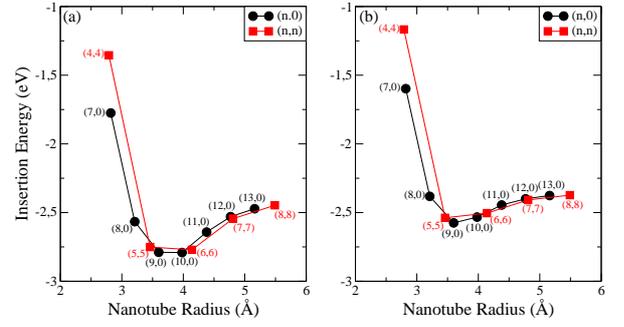}
\caption{\label{Figure2}(Color online) Insertion energy as a
function of the nanotube radius for the ground state of the Mo
wire encapsulated inside different zigzag and armchair nanotubes
without (a) and with (b) BSSE corrections, as discussed in the
text.}
\end{figure}

We have simulated Mo wires inside single-walled carbon nanotubes
with zigzag and armchair chiralities of different radii. All the
armchair nanotubes $(n,n)$ are metallic, while zigzag nanotubes
$(n,0)$ can behave as a metal or as an insulator, depending on
whether $n/3$ is an integer or not. Exhaustive calculations have
been performed by testing both dimerized and equidistant
encapsulated monoatomic Mo wires with different concentrations
(i.e. different separations between Mo dimers in the dimerized
chains and between atoms in the equidistant chain) inside the
nanotube. We always found the dimerized wires to be the most
stable ones for every nanotube. Among zigzag nanotubes, which have
a lattice constant of 4.29 \AA, the ground state was formed by a
dimerized Mo wire unit cell per unit cell of the (9,0) nanotube.
This meant that the length of the Mo wire was compressed a 6 \%
with respect to the free-standing configuration. Among armchair
nanotubes, which have a lattice constant of 2.48 \AA, the ground
state had a dimerized Mo wire unit cell per 2 unit cells of the
(5,5) nanotube, which led to a Mo wire stretched an 8 \% with
respect to the free-standing configuration. We illustrate in Fig.
(\ref{Figure1}) such configurations for the (9,0) zigzag nanotube
as well as for the (5,5) armchair one. Encapsulated dimerized Mo
wires with concentrations other than these ones, as well as
equidistant Mo wires, which were much higher in energy, are
revised at the end of this section.

\begin{figure}
\includegraphics[width=0.90\columnwidth]{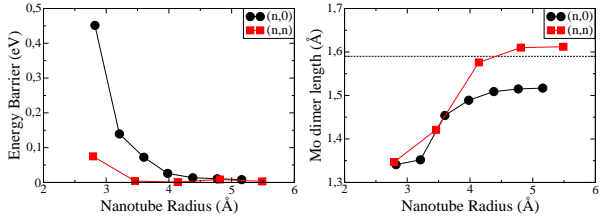}
\caption{\label{Figure3}(Color online) Diffusion energy barrier
(a) and length of the Mo dimer (b) as a function of the nanotube
radius for Mo wires encapsulated inside different zigzag and
armchair nanotubes.}
\end{figure}

We define the insertion energy of the Mo wire inside the nanotube
as the difference between the energy of the whole system
($E_{\mathrm{CNT}+n\mathrm{Mo}}$), and the energy of the isolated
carbon nanotube ($E_{\mathrm{CNT}}$) plus $n$ times the energy of
an isolated Mo atom ($n$ is the number of Mo atoms in the cell).
Finally, we divide by $n$ in order to normalize.

\begin{equation}
E_\mathrm{I}(\mathrm{Mo})=\frac{E_{\mathrm{CNT}+n\mathrm{Mo}}-E_\mathrm{CNT}-
n\,E_{\mathrm{Mo}}}{n}.
\end{equation}

Since we are using localized pseudoatomic orbitals in the basis
set, the calculation of the insertion energy must take into
account the basis set superposition error (BSSE).\cite{bsse} This
error arises due to the different dimension of the Hilbert spaces
associated to the whole system and each of its constituents taken
separately. To avoid this error, we have considered the ghost atom
method, which consists in including ghost atoms, i.e.,
pseudo-atomic orbitals without any atomic potential, in the
positions of the missing atoms to complete the Hilbert space. The
use of ghost atoms has been shown to be very efficient for
correcting the BSSE in other systems like, for instance, benzene
adsorbed on carbon nanotubes \cite{tournus} or pentacene
physisorbed on gold (001).\cite{lee} We plot in Fig.
(\ref{Figure2}) the insertion energy as a function of the nanotube
radius for different zigzag and armchair nanotubes with and
without the BSSE correction. We have found that the BSSE
correction has 2 effects in our calculations. On one hand, since
the energy of each separated component decreases when its Hilbert
space is enlarged, the absolute value of the insertion energy
decreases. On the other hand, the BSSE moves the radius for which
the minimum insertion energy is obtained to lower values.
Therefore, the BSSE correction is essential to produce correct
results, and all our calculations have taken it into account.

Fig. (\ref{Figure2})(b) and Table I show that the insertion energy mainly
depends on the diameter of the nanotube, and is independent of its
chirality. The thinnest nanotubes that were found to be able to
encapsulate a Mo wire were the (7,0) and the (4,4), both of which
have a radius around 2.8 \AA. Thinner nanotubes deform a lot upon
Mo insertion and lose their symmetry. The minimum in the insertion
energy was achieved for (9,0) and (5,5) nanotubes, with radii
around 3.5 \AA. For thicker nanotubes the bonding between the wire
and the nanotube slowly disappears and the insertion energy
increases asymptotically towards a constant value, due to the bonding
between Mo atoms. This asymptotic
limit corresponds to a Mo wire with interatomic distances close to
the free-standing case, i.e., without interactions with the
nanotube.

\begin{figure}
\includegraphics[width=0.90\columnwidth]{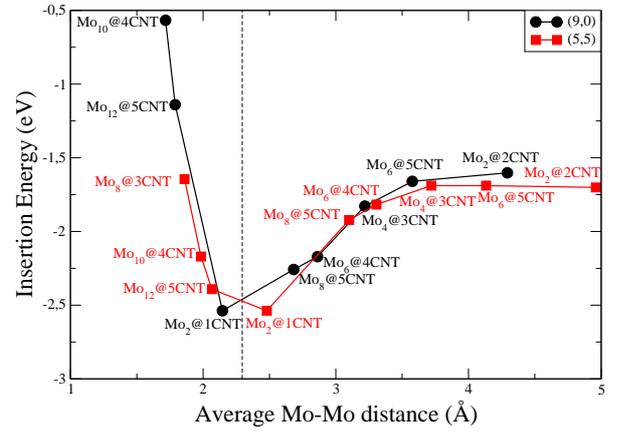}
\caption{\label{Figure4}(Color online) Insertion energy as a
function of the average distance between adjacent Mo atoms for
dimerized Mo wires of different concentrations encapsulated inside
(9,0) and (5,5) nanotubes. Each point Mo$_n$@mCNT corresponds to
$n$ Mo atoms inside $m$ carbon nanotube cells, where a zigzag cell
corresponds to a zigzag unit cell and an armchair cell corresponds
to 2 armchair unit cells. The vertical dashed line corresponds to
the free standing configuration.}
\end{figure}

The energy of the wire inside the carbon nanotube also varies as
the wire is moved along the axis of the nanotube. We define the
diffusion energy barrier as the maximum energy referred to the
ground state energy that the wire has to overcome when moving
along the nanotube, normalized per Mo atom. This energy barrier
has an essential importance in fixing the encapsulated wire, since
a too low energy barrier would allow the Mo atoms to move under
small forces or thermal agitation, possibly breaking the chain. Further,
the Mo atoms are expected to leave the chain just as liquids leave
a straw. To
obtain this energy barrier we have calculated the energy of the
system when the Mo wire is displaced along the carbon nanotube
from its relaxed structure, without relaxing again. In Fig.
(\ref{Figure3})(a), we plot the height of the energy barrier for
all of the previous systems. As expected, the value of the energy
barrier decreases rapidly when increasing the radius of the
nanotube, as the interaction between the nanotube and the wire
tends to disappear. We have surprisingly found that chirality in
this case plays a role, since armchair nanotubes present a much
lower energy barrier. This energy barrier becomes sizeable only
for armchair nanotubes with radii shorter than 3 \AA, as shown in
Fig. 3(a). Interestingly, although
the modulus of their insertion energy is about 1 eV lower than that of
the system with the optimal radius (see Fig. 2), the formation of the
structure with these short radii is still exothermic. Therefore, we expect
that encapsulated Mo chains could be seen experimentally in zigzag nanotubes
with radii of the order of 3 \AA.
It is also relevant to note that the energy
barrier in zigzag nanotubes is insensitive to their metallic or insulating
nature. It is not easy to deduce specific numbers for the atom mobility even
assuming an Arrhenius-type law for it, $\tau = \tau_0 exp(-E_{barrier}/K_B T)$,
since we do not know the typical time $\tau_0$ for activation.
 For a typical
experimental setup, we would use $\tau \sim $ few days $\sim 10^6$ seconds.
Introducing this
number, as well as a slow $\tau_0 \sim 10^{-10}$ seconds in
the equation above gives
temperatures for a working device
below $\sim 150K$. Hence, we predict that the experiments should be performed
at moderately low temperatures.

\begin{figure}
\includegraphics[width=0.90\columnwidth]{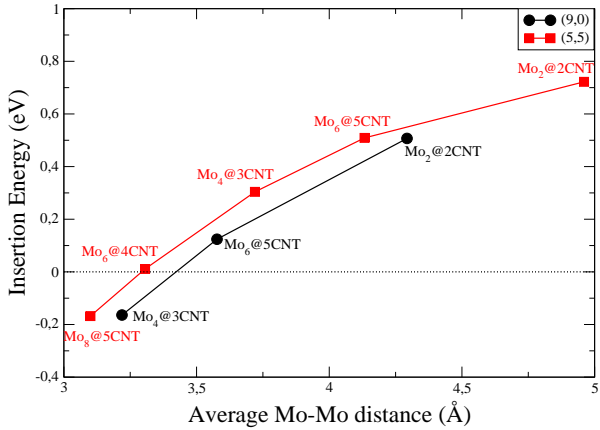}
\caption{\label{Figure10}(Color online) Insertion energy for
different equidistant Mo wires inside (9,0) and (5,5) carbon
nanotubes.}
\end{figure}

The structure of the chain is also affected by the nanotube. In
Fig. (\ref{Figure3})(b), we plot the length of the encapsulated Mo
dimers as a function of the nanotube radius, with the length of
the dimer in the free-standing Mo wire marked by a dashed line.
For short nanotube radii, the Mo dimers are compressed by the
effect of the nanotube potential. However, they enlarge towards a
constant value when the nanotube gets thicker. This constant value
for the dimer length is 1 \% longer for armchair nanotubes than in
the free-standing case, meanwhile inside zigzag nanotubes the
dimer length is compressed a 5 \%. The Mo dimer length does not
converge to the same value as in the free-standing case since we
impose a lattice constant condition to commensurate with the
lattice constant of the nanotube.


Dimerized Mo wires of different concentrations have also been
simulated inside (9,0) and (5,5) carbon nanotubes. We have studied
systems with a number $m$ of nanotube cells varying from 2 to 5
(where a zigzag cell corresponds to one zigzag unit cell and an
armchair cell corresponds to 2 armchair unit cells), and $n$ Mo
atoms in a dimerized structure. We plot in Fig. (\ref{Figure4})
the insertion energy of these arrangements, labeled as
Mo$_n$@$m$CNT, as a function of the average distance between
adjacent Mo atoms. The minimum in the insertion energy per Mo atom
is found for Mo$_2$@1CNT, for both (9,0) and (5,5) carbon
nanotubes. Then, Mo wires tend to resemble as much as possible the
free-standing one (dashed line). Increasing the concentration
leads to wires with an inter-dimer distance close to the dimer
length, which introduces a high repulsion and leads to a rapid
increase in the insertion energy. On the other hand, when the Mo
concentration is decreased, the insertion energy increases
asymptotically to the value of an isolated dimer inside the carbon
nanotube.

\begin{figure}
\includegraphics[width=0.90\columnwidth]{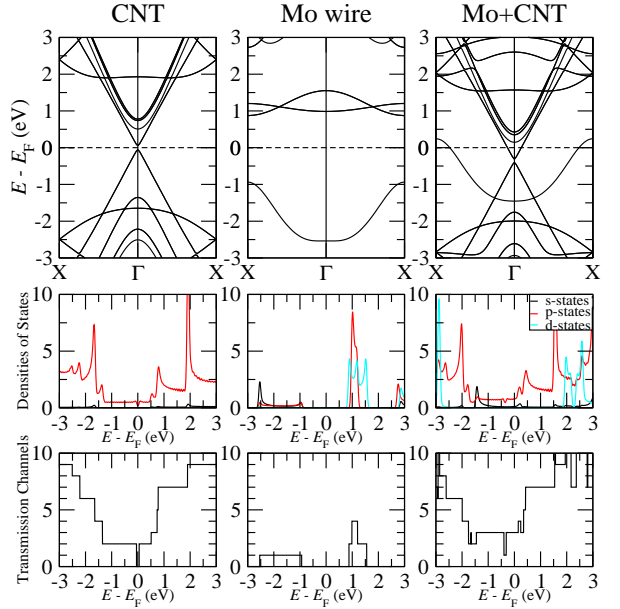}
\caption{\label{Figure5}(Color online) Band structure,
orbital-projected densities of states and transmission channels of
the system formed by a Mo dimerized wire encapsulated inside a
(9,0) carbon nanotube, compared with those of its constituents
separately.}
\end{figure}

Encapsulated equidistant Mo wires of the type Mo$_2$@1CNT, which
have a high Mo concentration, are not stable inside carbon
nanotubes since they tend to dimerize. However, if the Mo
concentration is decreased, the energy barriers that the nanotube
produces might make the system metastable. We have therefore
simulated equidistant Mo wires of different concentrations inside
(9,0) and (5,5) carbon nanotubes (Mo$_n$@$m$CNT), composed by $n$
equidistant Mo atoms inside $m$ nanotube cells, in a similar way
as we did with the dimerized wires. The results for the insertion
energy are shown in Fig. (\ref{Figure10}). The insertion energy
depends mainly on the distance between Mo atoms and is not much
affected by the type of nanotube, being a bit lower in the case of
zigzag nanotubes. In case of encapsulated
equidistant Mo wires with an interatomic distance between Mo atoms
shorter than around 3 \AA~we found that the Mo atoms tend to
dimerize and the wire is not able to retain its equidistant
arrangement. We also found that the insertion energy increases as
the distance between Mo atoms is increased, and becomes positive
for Mo-Mo distances larger than 3.25 \AA. Therefore, equidistant
Mo wires encapsulated inside carbon nanotubes could be achieved
with an interatomic distance between 3.00 \AA and 3.25 \AA, which
corresponds to the range where the insertion energy is negative.
However, these values are more than 2 eV higher than in the
dimerized arrangement.

Unlike the dimerized Mo wires, the equidistant wires are magnetic,
with an antiferromagnetic (AFM) coupling between adjacent Mo atoms
and a magnetic moment of around 5 $\mu$B per Mo atom. A
ferromagnetic (FM) state can also be found, with an energy
difference per Mo atom between the FM and the AFM which is larger
than 300 meV for the structures with negative insertion energy and
that asymptotically decreases towards zero as the distance between
Mo atoms increases.

\begin{figure}
\includegraphics[width=0.90\columnwidth]{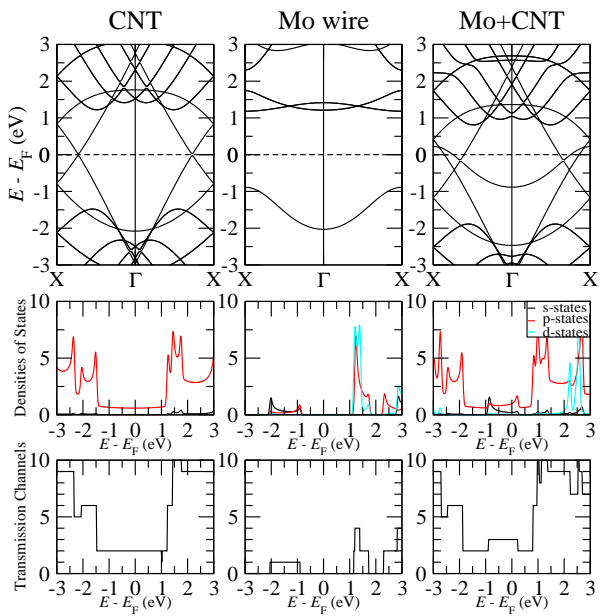}
\caption{\label{Figure6}(Color online) Band structure,
orbital-projected densities of states and transmission channels of
the system formed by a Mo dimerized wire encapsulated inside a
(5,5) carbon nanotube, compared with those of its constituents
separately.}
\end{figure}

\section{Electronic Structure and Transport Properties}

We have calculated the band structure, densities of states and
transmission channels of the Mo$_2$@1CNT for different thicknesses
and chiralities. Figs. (\ref{Figure5}) and (\ref{Figure6}) show
the results for the cases corresponding to the (9,0) and (5,5)
nanotubes, respectively. To clarify the discussion, we have also
calculated the above properties for the empty nanotube and the
isolated Mo wire with their geometries fixed to those within the
Mo$_2$@1CNT systems. The ground state of the free-standing Mo wire
exhibits a $sp_z$ band below the Fermi level ($z$ is the direction
of the wire) that leads to a transmission channel between -2 and
-1 eV.\cite{amador} This band is preserved when the wire is
compressed (Fig. (\ref{Figure5})) or stretched (Fig.
(\ref{Figure6})) inside the carbon nanotubes. However, it moves up
in energy as a consequence of the encapsulation, eventually
reaching the Fermi level. This leads to a low-bias total
conductance of 3 $\mathrm{G}_0$ due to 3 transmission channels at
the Fermi level, one from the $sp_z$ band of the Mo wire and two
others from the $p$ bands of the carbon nanotube.

\begin{figure}
\includegraphics[width=0.90\columnwidth]{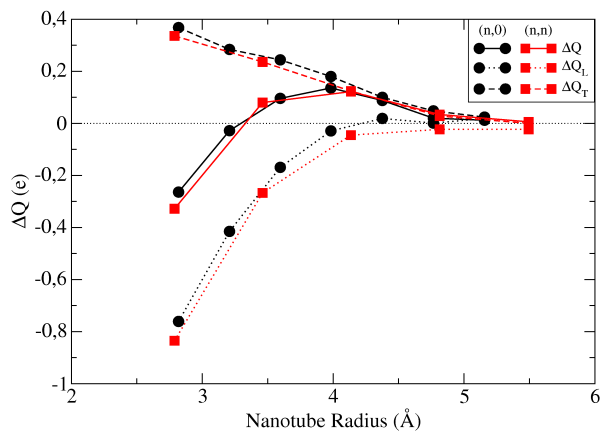}
\caption{\label{Figure7}(Color online) Charge variation per Mo
atom with respect to the Mo wire in the free-standing
configuration for the encapsulated Mo wire as a function of the
radius of the nanotube ($\Delta$Q, solid line). The charge
variation is decomposed in the longitudinal orbitals, $p_{x}$,
$p_{y}$, $d_{xy}$ and $d_{x^2-y^2}$ ($\Delta$Q$_\mathrm{T}$,,
dashed line) and the longitudinal orbitals, $s$ and $p_{z}$
($\Delta$Q$_\mathrm{L}$, dotted line).}
\end{figure}

\begin{figure}
\includegraphics[width=0.90\columnwidth]{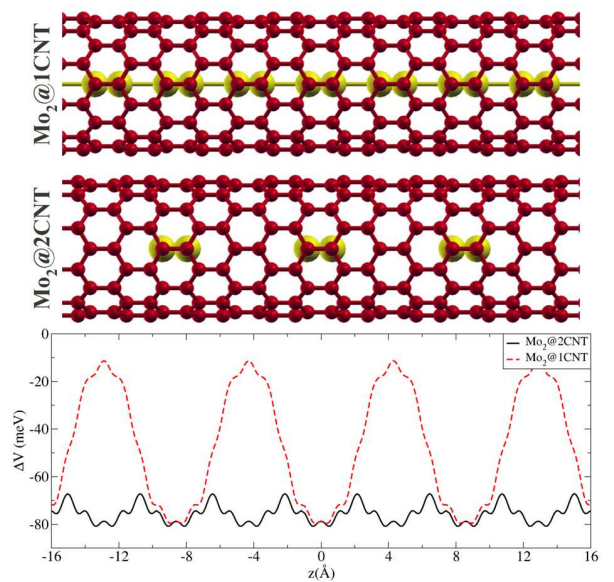}
\caption{\label{Figure9}(Color online) $\Delta V$ as a function of
the $z$ coordinate for the dimerized structures Mo$_2$@1CNT and
Mo$_2$@2CNT. The chirality of the nanotube is (10,0).}
\end{figure}

\begin{figure*}
\includegraphics[width=2\columnwidth]{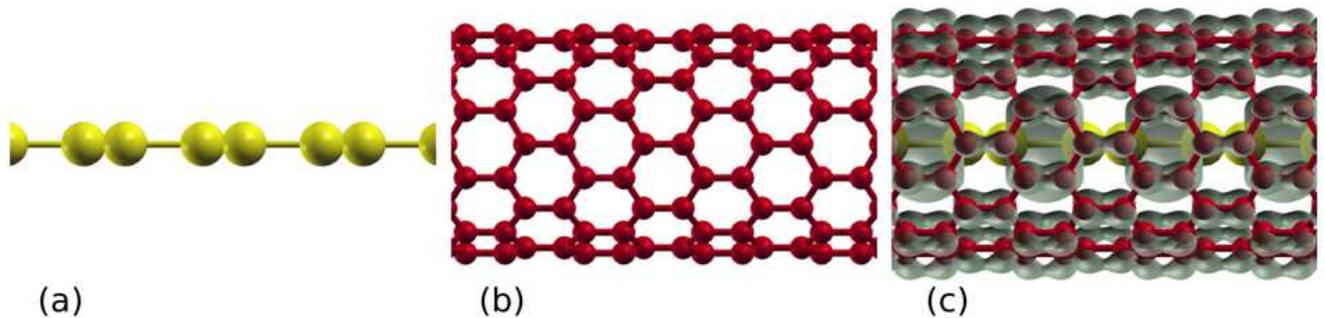}
\caption{\label{Figure8}(Color online) Charge density around the
Fermi level for the dimerized Mo wire (a) and the zigzag (10,0)
nanotube (b) (keeping fixed the structural configurations they
have in the hybrid nanostructure) and both of them in the hybrid
nanostructure (c).}
\end{figure*}

In order to analyze in more detail the bonding between the Mo wire
and the CNT, which is related to the displacement of some bands
and has important implications in the transport properties, we
have calculated the change in the electronic charge of the Mo
atoms encapsulated in the different nanotubes investigated. We
take as reference the charge in the ground state of the infinite
free-standing Mo wire. The results are shown in Fig.
(\ref{Figure7}). We note that, inside very thin nanotubes, Mo
tends to lose charge, meanwhile the charge transfer changes its
sign when the nanotube radius increases, leading to a maximum for
(10,0) and (6,6) nanotubes. For nanotubes with large enough radii,
the charge transfer tends to 0 as the interaction between the wire
and the nanotube disappears. Therefore, Mo wires inside (9,0) and
(5,5) nanotubes, which correspond to the most stable
configurations, increase their charge. We have also calculated the
charge variation in each cartesian orbital. We have found that the
$d_{xz}$, $d_{yz}$ and $d_{3z^2-r^2}$ orbitals are only slightly
affected by the encapsulation; $p_{x}$, $p_{y}$, $d_{xy}$ and
$d_{x^2-y^2}$ (which are transversal orbitals) increase their
charge, and $s$ and $p_{z}$ (which are longitudinal orbitals) lose
charge, which is consistent with the $sp_z$ band crossing the
Fermi level (see Figs. (\ref{Figure5}) and (\ref{Figure6})). Fig.
(\ref{Figure7}) also shows the charge variation in the transversal
and longitudinal orbitals separately. The charge transfer towards
Mo transversal orbitals is favored by the encapsulation, decreases
linearly as the nanotube radius increases and is still present for
relatively large nanotube radii. On the other hand, Mo
longitudinal orbitals lose charge for small nanotube sizes, but
this loss  rapidly disappears (exponentially) when the nanotube
radius increases. The excess of charge in the transversal orbitals
and its disappearance for large CNT radii can be explained as an
effect of the bonding between these orbitals and the $p$ orbitals
from the C atoms, which is reflected in the strong hybridization
between them.

We have also analyzed the change in the electrostatic potential at
the nanotube walls when the Mo wire is introduced \cite{Victor3}.
Such potential could be able to trap electrons,\cite{Lee02,Cho03}
produce entangled electrons pairs and tailor the interaction
between static and flying qubits \cite{Gun06}. To estimate this
change, which is related to charge transfer, structural
modification and hybridization between the Mo atoms and the
nanotube, we have calculated the electrostatic potential at the
walls of the nanotubes with and without the Mo wire and subtracted
one from the other, i.e. $\Delta
V(z)=V_\mathrm{with}(z)-V_\mathrm{without}(z)$. We note that this
potential only depends on the $z$ coordinate due to the axial
symmetry. We plot in Fig. (\ref{Figure9}) $\Delta V(z)$ for 2
dimerized Mo wires with different concentrations inside a (10,0)
carbon nanotube. As expected, clearly defined wells appear around
the dimers, with a valley in the position of each Mo atom due to
the presence of more negative charge at these points. The absolute
maximum in $\Delta V(z)$ appears at the middle point between
dimers. For Mo$_2$@2CNT, $\Delta V(z)$ has enough room to reach
nearly  zero, meanwhile for Mo$_2$@1CNT this is not possible.

The displacement of the $sp_z$ band from the Mo wire towards the
Fermi level is concomitant with the charge variation of these
states, as illustrated in Fig. (\ref{Figure7}) and discussed
above. Therefore, regardless of the chirality of the system, if
the nanotube radius is small enough to produce a noticeable charge
variation in the longitudinal orbitals, the $sp_z$ band of the Mo
wire will enter the Fermi level and contribute to the conduction
(see Figs. (\ref{Figure5}) and (\ref{Figure6})). By calculating
the band structure and transmission channels of each structure, we
have found that the $sp_z$ band enters the Fermi level for zigzag
nanotubes up to the (10,0) and for armchair nanotubes up to the
(6,6), which means that there is a transition radius around 4.2
\AA. This implies that, if the nanotube has an appropriate radius,
when the insulator dimerized Mo wire is encapsulated, the
resulting system has metallic behavior, even if the nanotube is
also an insulator.  As an example, we plot in Fig. (\ref{Figure8})
the density of states in real space integrated around the Fermi
level for the dimerized Mo wire inside the zigzag (10,0) nanotube,
and we compare it with the corresponding density of states of its
isolated constituents. Since both the isolated nanotube and Mo
wire are insulators, there is no density of states around the
atoms as illustrated in the picture. However, when the hybrid
quasi one-dimensional system is formed, a noticeable amount of
density of states is present. This density of states is
delocalized through the whole system (both C atoms and Mo atoms)
and results from the hybridization between the $sp_z$ band of Mo
with the $p$ bands of C around the Fermi level. These states make
the hybrid system conductive at low bias voltage. Therefore,
encapsulating Mo wires in CNT is a way to create conductive quasi
one-dimensional hybrid nanostructures from their insulating
constituents.

\section{Conclusions}

We have investigated monoatomic chains of molybdenum encapsulated
in single walled carbon nanotubes of different chiralities. The
electronic structure and the optimal atomic arrangement have been
calculated using the DFT code SIESTA with the generalized gradient
approximation. We have also used the non-equilibrium Green's
functions formalism as implemented in the SMEAGOL code
\cite{smeagol} to calculate the conductance in the ohmic
regime.

We have found that the insertion energy mainly depends on the
diameter of the nanotube, and is independent of its chirality. The
thinnest nanotubes able to encapsulate a Mo wire were the (7,0)
and the (4,4), both of them with a radius around 2.8 \AA. Among
zigzag nanotubes, which have a lattice constant of 4.29 \AA, the
ground state was formed by a dimerized Mo wire unit cell per unit
cell of the (9,0) nanotube. This meant that the length of the Mo
wire was compressed a 6 \% with respect to the free-standing
configuration. Among armchair nanotubes, which have a lattice
constant of 2.48 \AA, the ground state had a dimerized Mo wire
unit cell per 2 unit cells of the (5,5) nanotube, which led to a
Mo wire stretched an 8 \%. These nanotubes, which are the optimal
for encapsulating Mo wires, have radii around 3.5 \AA.

We have also calculated the diffusion energy barrier, which
decreases rapidly when the radius of the nanotube is increased. We
have demonstrated that chirality in this case plays a role, since
armchair nanotubes present a lower energy barrier. A too low
diffusion energy barrier could hamper however the experimental
production of such structures since the wire would be able to move
easily along the nanotube under mechanical or thermal
perturbations.

Finally, we have shown that the bonding between the transversal
orbitals of Mo and the $p$ orbitals of the C atoms upon
encapsulation leads to electronic charge transfer towards those Mo
transversal orbitals. We have found that this transfer is still
present for relatively large nanotube radii. The $sp_z$ band of
the insulator free-standing dimerized Mo wire was preserved when
the wire was compressed or stretched inside the carbon nanotubes
but it moved up in energy as a consequence of the encapsulation,
eventually reaching the Fermi level and contributing with the $p$
bands of the CNT to a low-bias total conductance of 3
$\mathrm{G}_0$. The $sp_z$ band crossed the Fermi level for zigzag
nanotubes up to the (10,0) and for armchair nanotubes up to the
(6,6). This implies that, if the nanotube has an appropriate
radius, like the (9,0) or the (5,5), when the insulator dimerized
Mo wire is encapsulated the resulting system has always metallic
behavior, even if the nanotube is also an insulator. Therefore,
encapsulating Mo wires in CNTs can lead to metallic nanostructures
comprised of parts which are insulating on their own.

\acknowledgments This work was supported by the Marie Curie network NanoCTM,
by the Spanish Ministry
of Education and Science in conjunction with the European Regional
Development Fund (Projects FIS2008-02490/FIS and FIS2009-07081),
and by Junta de Castilla y Le\'on (Project GR120). VMGS thanks the
Spanish Ministerio de Ciencia e Innovaci\'on for a Juan de la Cierva fellowship.

{}

\end {document}